# Interpretation of experimental evidence of the topological Hall effect.

A. Gerber

Raymond and Beverly Sackler Faculty of Exact Sciences,

School of Physics and Astronomy, Tel Aviv University, Tel Aviv 6997801, Israel

Topological Hall effect in magnetic materials is considered the ultimate trademark of the skyrmion phase. The phenomenon is identified by distinct non-monotonic features in the Hall effect signal presumed to be the evidence of the topological origin. It is demonstrated here that similar features, unrelated to the skyrmion physics, arise in heterogeneous ferromagnets when components of the material exhibit the extraordinary Hall effect with opposite polarities. Relevance of this mechanism to the published data is discussed.

Extraordinary or anomalous Hall effect (EHE) in ferromagnetic materials is a well-known phenomenon discovered more than a century ago. In a vast majority of the studied materials, the EHE signal is proportional to magnetization, which makes the EHE one of the standard magnetometric techniques. However, a linear correlation between magnetization and the EHE does not hold in heterogeneous ferromagnetic systems where the Hall coefficient is not uniform. Unusual features appearing in such cases deserve attention, in particular when searching for novel phenomena like the topological Hall effect.

Magnetic skyrmions is a fashionable subject inspired by their fundamentally non-trivial topological origins and potential applications as bits of information in future memory and logic devices. Stable ground-state skyrmions were predicted [1] to form in materials lacking inversion symmetry due to a non-centrosymmetric crystal lattice structure [2, 3] or due to antisymmetric exchange interactions that occur near the symmetry breaking magnetic interfaces [4]. The skyrmion phase can be observed by Lorentz transmission electron microscopy [5-7], magnetic force microscopy [8], Kerr microscopy [9, 10], spin-polarized scanning tunneling microscopy (SPSTM) [11, 12], spin-polarized low energy electron microscopy (SPLEEM) [13] and neutron scattering [14-16]. However, the largest share of the reported experimental evidence is based on the topological Hall effect. When a conduction electron passes through a skyrmion, its spin experiences a fictitious in real space magnetic field, which deflects the conduction electrons perpendicular to the current direction. The phenomenon, termed the topological Hall effect (THE) [17, 18], can be observed as a distinct, additional contribution in Hall measurements superposed on the ordinary and the extraordinary Hall effects. Such a distinct feature has been found in the A phase of MnSi [19, 20] consistently with an observation of the skyrmion lattice by neutron scattering [14]. Since then, the THE was accepted as a hallmark of topologically nontrivial (chiral) spin textures and multiple later works used observation of the THE features as a sufficient evidence of the skyrmion phase [21 – 32]. It is demonstrated in the following that the features attributed to the THE can be generated by the extraordinary Hall effect in heterogeneous ferromagnets without involving skyrmions.

In papers dealing with the topological Hall effect, the Hall resistivity is presented as:

$$\rho_{xy}(B) = R_{OHE}B + \mu_0 R_{EHE} M(B) + \rho_{THE} \tag{1}$$

where the first term is the ordinary Hall effect with $R_{OHE}$ being the ordinary Hall effect coefficient and $B$ the magnetic induction, the second term is the extraordinary or anomalous Hall effect (EHE) term with $R_{EHE}$ being the extraordinary Hall effect coefficient and M the normal to plane magnetization, and $\rho_{THE}$ is the topological Hall effect term. The ordinary effect is taken as a linear function of the applied field and is usually neglected (we don't discuss the non-linear cases where two or more charge carriers are present). Coefficient $R_{EHE}$ is assumed constant. Magnetization $M$ is a monotonically increasing function of the applied field up to magnetic saturation. Therefore, the EHE term is expected to be a smooth monotonically increasing function of the applied field until saturation at high field. Observation of anomalies in the Hall resistivity, sometimes in the form of pronounced non-monotonic in field bumps, is taken as the evidence of the topological Hall effect. $\rho_{THE}$ is then determined as:

$$\rho_{THE} = \rho_{xy}(B) - \mu_0 R_{EHE} M(B) \tag{2}$$

where the EHE term is estimated using the measured magnetization $M(B)$ or by a smooth and monotonic extrapolation to the saturated high field value. The assumption of a linear correlation between the EHE and magnetization can be erroneous if the material is not homogeneous.

Let's assume a heterogeneous system composed of two parallel magnetically decoupled ferromagnetic layers, each exhibiting its own magnetization with the corresponding hysteresis, the coercive and the saturation fields and the respective extraordinary Hall effect. For simplicity, let's assume that layers have comparable resistance. The total EHE voltage is a superposition of two parallel signals generated in each layer separately, approximated as:

$$V_{EHE} \approx \frac{1}{2}\left(V_{EHE,1} + V_{EHE,2}\right) \approx \frac{1}{4}\mu_0 I \left(\frac{R_{EHE,1} M_1}{t_1} + \frac{R_{EHE,2} M_2}{t_2}\right) \tag{3}$$

where $I$ is current equally split between the two layers, and $M_i$, $R_{EHE,i}$ and $t_i$ are the respective magnetization, EHE coefficient and thickness of each layer. Magnetization per square of such system would be:

$$M = \frac{M_1 t_1 + M_2 t_2}{t_1 + t_2} \qquad (4)$$

The EHE signal (Eq.3) is not proportional to magnetization (Eq.4) if $R_{EHE,1} \neq R_{EHE,2}$. The difference can be qualitative when the EHE coefficients of two layers have opposite polarities. In this case, the field dependence of the observed signal can become non-monotonic. Fig.1a presents the EHE voltage hysteresis loops of two individual ferromagnetic layers, the first with a positive EHE coefficient $R_{EHE}^+$ and the second with a negative one $R_{EHE}^-$. $V_{EHE,sat}^+$ and $V_{EHE,sat}^-$ are the saturated EHE voltages generated at high positive field in the layers with respectively positive $R_{EHE}^+$ and negative $R_{EHE}^-$. The coercive field of the layer with negative $R_{EHE}^-$ is larger than that of the positive one: $B_c^- > B_c^+$. Superposition of two signals is shown in Fig.1b. The total EHE signal of such two-layer system is a non-monotonic function of applied field with a characteristic bump feature. The bump develops in the field range $B_c^+ \leq B \leq B_c^-$. The saturated EHE voltage at high positive field is negative when: $|V_{EHE,sat}^+| < |V_{EHE,sat}^-|$ and positive when $|V_{EHE,sat}^+| > |V_{EHE,sat}^-|$. The sketch presents both layers exhibiting hysteresis with different coercive and saturation fields. Obviously, a non-monotonic signal will develop also in absence of hysteresis if the EHE polarity of the two layers are opposite and their saturation fields are different.

Experimental demonstration of a variety of cases can be found in Co/Pd bilayers and multilayers. Co and Pd are completely soluble and form an equilibrium fcc solid solution phase at all compositions [33]. The EHE polarity of the solution reverses at room temperature at Co atomic concentration of about 38% from positive in Co-rich alloys to negative in Pd-rich ones [34, 35]. In Co/Pd bilayers and multilayers an inter-diffusion between the two components takes place at the interfaces, thus forming a material with spatially and temporarily varying composition. The process of inter-diffusion and alloying of the interfaces occurs as a natural aging at room temperature or can be accelerated by annealing [34]. The EHE coefficient of Co is positive while that of the Pd-rich alloyed interface is negative. Fig.2 presents an example of the Hall resistivity of Co/Pd bilayer film measured shortly after the fabrication (open rhombs) and half year later (solid circles). The sample was produced by rf sputtering from two targets. Thickness of Pd and Co layers are 5 nm and 2 nm respectively. The EHE signal of the fresh sample is that of Co, which is positive, monotonic and proportional to the film magnetization. The signal of the aged sample is non-monotonic, composed of a positive contribution of cobalt and a negative contribution from the interface CoPd alloy. The saturation field of the interface alloy is lower than that of cobalt, resulting in an unusual non-monotonic signal. After a complete and homogeneous inter-mixing, the signal becomes monotonic with the final polarity depending on the ratio between the initial amounts of Co and Pd [34]. This reversible and non-monotonic in field EHE signal is similar to the pattern attributed to the topological Hall effect in e.g. $SrRuO_3/LaSrMnO_3$ bilayers [23] and $Mn_3Ga$ [24].

Superposition of the reversible and irreversible EHE contributions with opposite polarities is shown in Fig.3. The sample is $[Co_{0.36} / Pd_{0.62}]_{20}$ multilayer produced by consecutive sputtering with thickness given in nm. Strained alloyed interfaces in Co/Pd multilayers give rise to the perpendicular magnetic anisotropy and hysteresis in the field loops [34, 36]. When Co layers are relatively thick and the intermixing is heterogeneous, the EHE can get the non-monotonic form shown in Fig. 3. The signal is composed of two components: hysteresis with a negative EHE coefficient contributed by the alloyed interfaces, and the reversible positive EHE term contributed by the internal Co or Co-rich alloy. The saturation field of the hysteresis component is lower than that of cobalt, resulting in a peculiar non-monotonic in field EHE loop with hysteresis. The signal is similar to that observed in e.g. FeGe [25, 26] and MnGa/heavy metal bilayers [27] and interpreted as the topological Hall effect.

Superposition of two irreversible EHE signals with opposite polarities is shown in Fig. 4. The sample is a two-level multilayer structure $[Co_{0.2}/Pd_{0.9}]_6 / [Co_{0.4}/Pd_{0.9}]_6$. Both multilayers exhibit perpendicular anisotropy with different coercive fields, while the EHE coefficient of $[Co_{0.2}/Pd_{0.9}]_6$ is negative and that of $[Co_{0.4}/Pd_{0.9}]_6$ is positive. The resulting signal is an experimental implementation of the model case sketched in Fig.1 with $B_c^- >$ $B_c^+$ and $|V_{EHE,sat}^+| < |V_{EHE,sat}^-|$. The magnitude and width of the bump feature can be modified artificially by adjusting the coercive field and the relative magnitude of the EHE contributions from each multilayer component via the relative thickness of Co and Pd layers, the repetition number, thickness of each strata and temperature. Such multilayer structures were proposed to serve as the multibit EHE magnetic random access memory units [37]. Signals similar to the one in Fig.4 were attributed to the topological Hall effect in SrRuO-SrIrO bilayers [28], SrRuO/SrIrO/SrRuO trilayers [29], Mn-doped $Bi_2Te_3$ [30]; $Mn_2CoAl$ capped by Pd [31] and heterostructures Cr(BiSb)Te/(BiSb)Te [32].

Reversal of the EHE polarity with composition and temperature is not restricted to Co/Pd and was found in multiple materials. In a number of cases, development of the unusual bump features attributed to the topological Hall effect was observed in a limited range of temperature, electric field and structure where the saturated EHE signal reversed its polarity between positive and negative. Bumps were observed in the EHE polarity reversal range of temperature in $Mn_2CoAl$ capped by Pd [31], $SrRuO_3$/LaSrMnO$_3$ bilayers [23], EuO [22] and heterostructures Cr(BiSb)Te/(BiSb)Te [32]; in the reversal range of composition in MnGa/Pt [27]; and in the temperature and electric field range in SrRuO/SrIrO heterostructures [29]. It might be plausible in some cases that the material is not homogeneous within this limited composition, temperature or electric field range, but contains two phases with opposite EHE polarity, and the anomalous non-monotonic pattern is a result of superposition of the two EHE contributions.

To summarize, the extraordinary Hall effect in heterogeneous ferromagnetic systems is generally not proportional to the material's magnetization. The differences are particularly visible when polarity of the EHE coefficient in different ferromagnetic regions are opposite. The field dependence of the extraordinary Hall effect can be non-monotonic with unusual features similar to those attributed to the topological Hall effect. As the conclusion, non-monotonicity of the observed Hall signal should not be taken as an unambiguous signature and sufficient evidence of the topological Hall effect without additional testing.

I acknowledge Amir Segal and Gal Winer for Co/Pd data used in this paper. The work was supported by the Israel Science Foundation grant No. 992/17 and the State of Israel Ministry of Science, Technology and Space grant No.53453.

**Figure captions.**

Fig.1. (a) EHE voltage hysteresis loops of two ferromagnetic layers, the first with a positive EHE coefficient $R_{EHE}^+$ (blue line online) and the second with a negative one $R_{EHE}^-$ (red line online). $V_{EHE,sat}^+$ and $V_{EHE,sat}^-$ are the saturated EHE voltages generated at high positive field in the layers with respectively positive $R_{EHE}^+$ and negative $R_{EHE}^-$. $|V_{EHE,sat}^+| < |V_{EHE,sat}^-|$. The coercive field of the layer with negative $R_{EHE}^-$ is larger than that of the positive one: $B_c^- > B_c^+$. Arrows indicate direction of the field sweep.

(b) Superposition of the two signals.

Fig.2. Hall resistivity of $Pd_5$ / $Co_2$ bilayer sample shortly after the deposition (open rhombes) and half year later (solid circles). Thickness of Pd and Co layers are 5 nm and 2 nm respectively.

Fig. 3. Hall resistance of $[Co_{0.36} / Pd_{0.62}]_{20}$ multilayer sample as a function of normal to plane field. Thickness is in nm.

Fig. 4. Hall voltage of a two-level multilayer structure $[Co_{0.2}/Pd_{0.9}]_6/[Co_{0.4}/Pd_{0.9}]_6$ as a function of normal to plane field. The EHE coefficient of $[Co_{0.2}/Pd_{0.9}]_6$ is negative and that of $[Co_{0.4}/Pd_{0.9}]_6$ is positive.

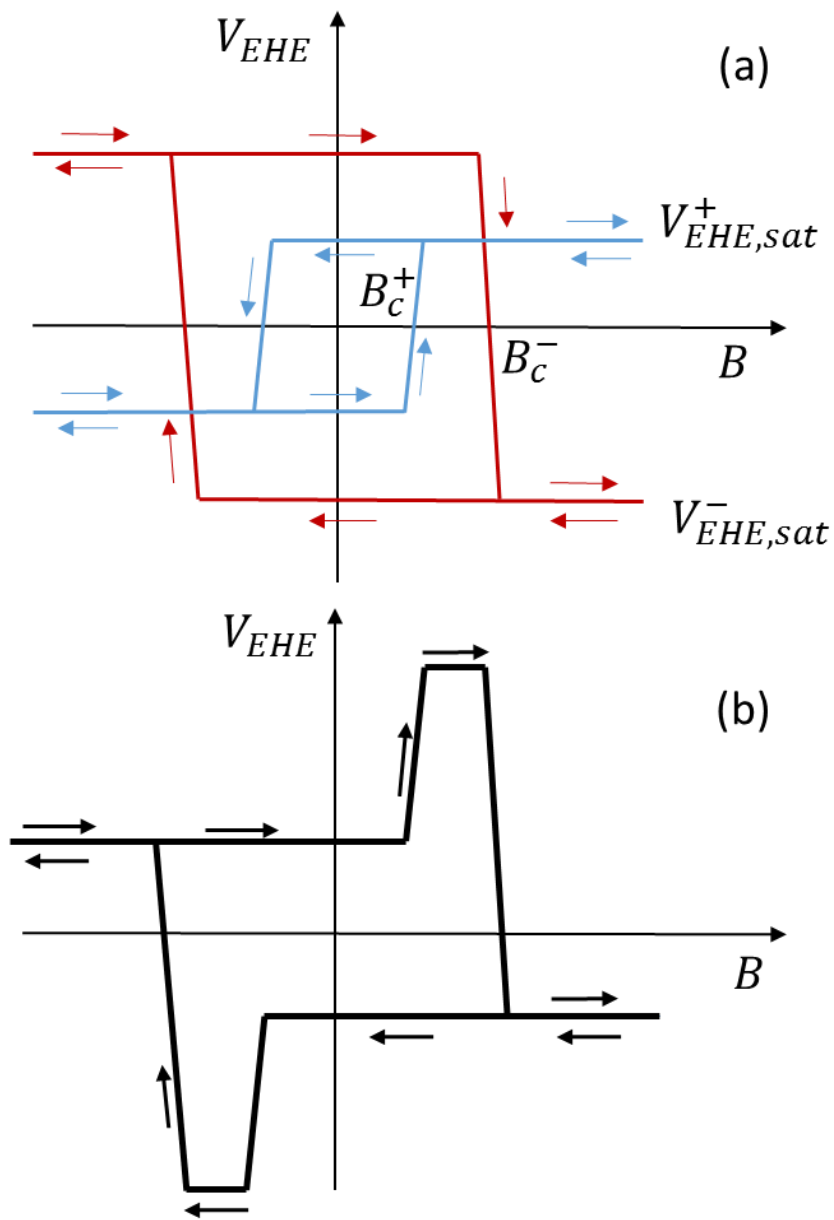

Fig. 1

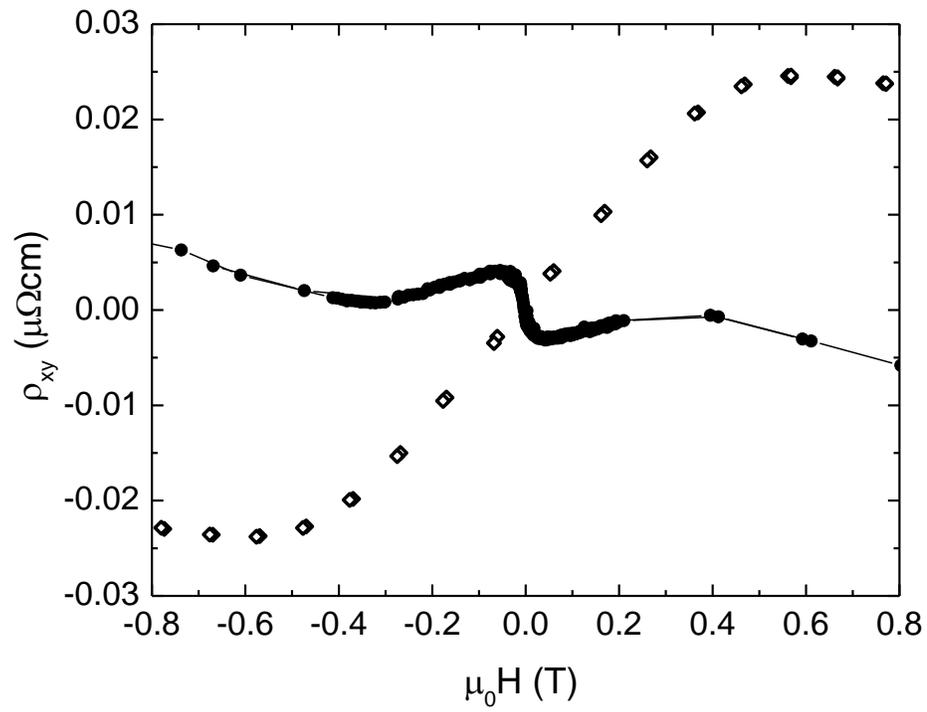

Fig. 2

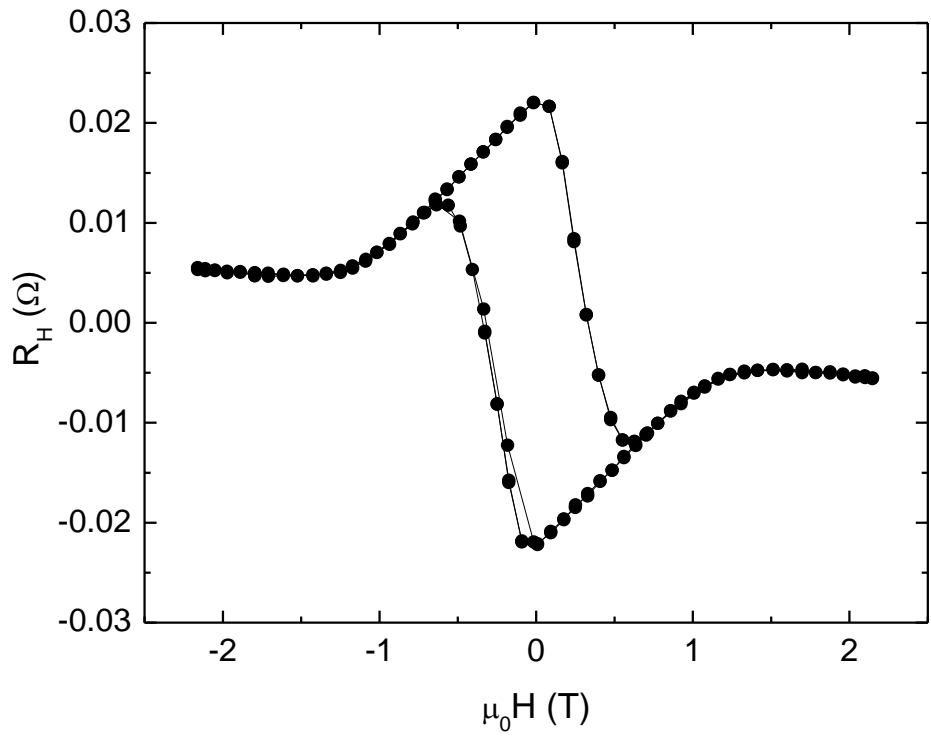

Fig. 3

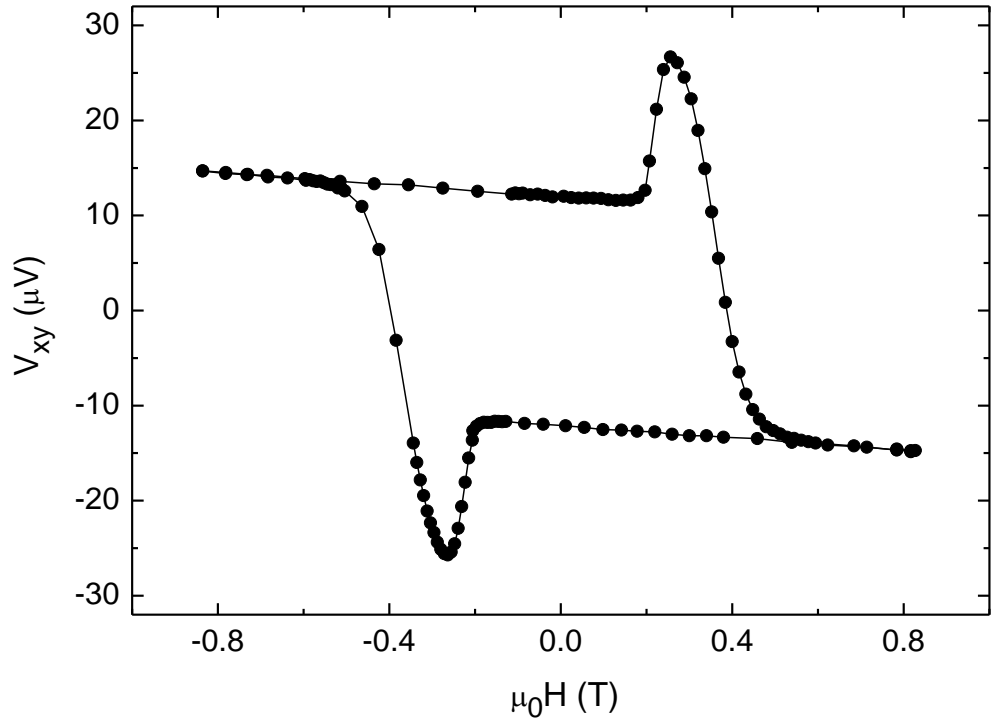

Fig. 4